\documentclass[aps,pre,reprint,showpacs,superscriptaddress，twocolumn]{revtex4-1}
\usepackage{graphicx}
\usepackage{amsmath}
\usepackage{multirow}
\begin{document}


\title{Uncovering Universal Wave Fluctuations In a Scaled Ray-Chaotic Cavity With Remote Injection}


\author{Bo Xiao}
\affiliation{Department of Electrical and Computer Engineering, University of Maryland, College Park, Maryland 20742-3285, USA}
\affiliation{Center for Nanophysics and Advanced Material, Physics Department, University of Maryland, College Park, Maryland 20742-3285, USA}
\author{Thomas M. Antonsen}
\author{Edward Ott}
\affiliation{Department of Electrical and Computer Engineering, University of Maryland, College Park, Maryland 20742-3285, USA}
\affiliation{Center for Nanophysics and Advanced Material, Physics Department, University of Maryland, College Park, Maryland 20742-3285, USA}
\author{Zachary B. Drikas}
\author{Jesus Gil Gil}
\affiliation{U.S. Naval Research Laboratory, Washington, DC, USA}
\author{Steven M. Anlage}
\affiliation{Department of Electrical and Computer Engineering, University of Maryland, College Park, Maryland 20742-3285, USA}
\affiliation{Center for Nanophysics and Advanced Material, Physics Department, University of Maryland, College Park, Maryland 20742-3285, USA}

\date{\today}

\begin{abstract}
The Random Coupling Model (RCM), introduced by Zheng, Antonsen and Ott \cite{zheng1,zheng2}, predicts the statistical properties of waves inside a ray-chaotic enclosure in the semi-classical regime by using Random Matrix Theory, combined with system-specific information. Experiments on single cavities are in general agreement with the predictions of the RCM. It is now desired to test the RCM on more complex structures, such as a cascade or network of coupled cavities, that represent realistic situations, but which are difficult to test due to the large size of the structures of interest. This paper presents a novel experimental setup that replaces a cubic-meter-scale microwave cavity with a miniaturized cavity, scaled down by a factor of 20 in each dimension, operated at a frequency scaled up by a factor of 20 and having wall conductivity appropriately scaled up by a factor of 20. We demonstrate experimentally that the miniaturized cavity maintains the statistical wave properties of the larger cavity. This scaled setup opens the opportunity to study wave properties in large structures such as the floor of an office building, a ship, or an aircraft, in a controlled laboratory setting.
\end{abstract}

\pacs{}

\maketitle

\section{Introduction}
Consider a partially open complex electrically-large enclosure being subjected to an incoming electromagnetic wave. A common problem of interest for electromagnetic compatibility and telecommunications is that of finding the induced voltage on an object at an arbitrary location inside the enclosure. Complex enclosures, such as computer cases with circuitry inside, or offices filled with desks, chairs, and electronics, are examples of ray-chaotic systems. To define what we mean by ray-chaotic, consider the case where the wavelength is short, two rays starting from the same location in such an enclosure but with slightly different directions. As the rays propagate reflecting from either curved surfaces or the interior features of the enclosure, their separation will tend to increase exponentially in time, and we call such situations ray-chaotic. Ray chaos leads to an extreme sensitivity to initial conditions for the rays \cite{ottBook}. For waves propagating in highly over-moded ray-chaotic structures, the exact solution for the fields depends  strongly  on the geometric details of the structure and is very sensitive to small changes in frequency or geometry. Thus, in the presence of even small uncertainties in structure or frequency, a statistical approach may be more appropriate than trying to obtain an exact solution for field quantities inside the structure \cite{Holland1999}. The Random Coupling Model (RCM) is one such method to predict the statistical properties of the waves inside a ray-chaotic enclosure \cite{zheng1,zheng2}. The RCM has been widely discussed and tested over the years, with good agreement between theory and experimental results on individual complex structures \cite{sameerRCM,Gradoni2014606,PhysRevE.86.046204,sameerthesis,zachgigabox,bisrat}.

There is interest in using the RCM  to understand the wave properties of more complex structures, such as a cascade or a network of coupled cavities. It becomes increasingly difficult to experimentally test these structures due to their large size and the difficulty in managing and reconfiguring them in a typical laboratory environment. To solve this problem, we propose miniaturizing the complex structure while maintaining the statistical properties of the waves by carefully scaling the frequency and the quality factor of the system. Electromagnetic geometric scale modeling has been used extensively in simulations and modeling of large structures for decades \cite{1697563,EMScaleNote}. The idea of scaling down the geometric size is not new in modeling, but the challenge is to make other electromagnetic properties scale appropriately as well. In this paper, we demonstrate the process by scaling down in size a cubic meter box, which is well studied in \cite{zachgigabox,GilGil}, and we experimentally demonstrate that the appropriately miniaturized enclosure has electromagnetic properties that are statistically identical to the full-scale enclosure. A key point in our scaling implementation is that, along with the straightforward scaling of size and frequency, it is also crucial to appropriately scale the conductivity of metal structure. This sets the stage for future investigations of complex structures. As part of this process we also demonstrate that a wave chaotic enclosure can be interrogated remotely to assess and fully characterize its statistical properties.

\section{Random Coupling Model (RCM)}
\label{sec:alphaTheory}
The RCM is based on Random Matrix Theory (RMT), originally proposed to model the energy level statistics of heavy nuclei \cite{wigner}. The idea is that if the wave system is sufficiently complex then its appropriate statistical properties are the same as those of a suitable ensemble of random matrices. Certain statistical properties, such as the distribution of the normalized spacings between nearest neighbor eigenfrequencies, follow a universal behavior regardless of the system details. It is difficult to identify these universal statistical properties in experimentally measured data because it inevitably contains system-specific features like the coupling between the ports and the cavity modes and short orbits \cite{PhysRevE.80.041109,PhysRevE.93.052205,PhysRevE.82.041114}. The RCM introduces a framework to incorporate the non-universal features with the universal statistical properties of appropriate random matrices to reproduce in the statistical sense the experimentally measured cavity impedance matrices. The effect of uniformly distributed loss in the system is a sub-unitary scattering system \cite{PhysRevB.55.4695}, and this effect is captured to very good approximation by a single loss parameter $\alpha$ \cite{zheng1,zheng2}. The RCM is formulated in terms of the impedance matrix $\mathbf{Z}$ of an $N$-port system. The ports represent sources or sinks of radiation that introduce or absorb energy in the enclosure. The impedance relates the voltage induced on one port to the currents at all of the $N$ ports, and is simply related to the $N\times N$ scattering matrix $\mathbf{S}$ through a bilinear transformation $\mathbf{S}=\mathbf{Z}_0^{1/2}(\mathbf{Z}+\mathbf{Z}_0)^{-1}(\mathbf{Z}-\mathbf{Z}_0)\mathbf{Z}_0^{-1/2}$ is a diagonal real matrix whose elements are the characteristic impedance of the transmission line modes connected to each port.

The loss parameter $\alpha$ is the ratio of the typical 3-dB bandwidth of the resonance divided by the mean spacing between modes. For a given system, $\alpha$ can be obtained in different ways, depending on what is known about the cavity. If the volume, $V$, and the typical quality factor, $Q$, are known, then $\alpha$ can be computed directly from its definition $\alpha=k^3V/(2\pi^2Q)$, where $k$ is the wave number. Otherwise, one can adopt the RCM normalization prescription, \cite{zachgigabox,sameerRCM} (summarized in the appendix), which estimates $\alpha$ by fitting the RCM prediction to the measured probability distribution functions of the cavity impedance.

\section{Scaling of The Cavity}
Our objective is to take a full scale complex enclosure of volume $V\approx 1\textnormal{ m}^3$ and create a scaled-down-in-size version with the same statistical electromagnetic properties. To reduce the cavity linear scale by a factor of 20 and increase the frequency commensurately is straight forward. However, the challenge is to maintain the same loss parameter ($\alpha$) value (hence the same statistical properties). If a cavity of volume $V$ is scaled down by a factor of $s$ in each dimension, giving a new volume of $V'=V/s^3$, then the wavelength and wavenumber scale as $\lambda'=\lambda/s$ and $k'=ks$. Experimentally, frequency scaling can be achieved by using frequency extenders, which are frequency multipliers that convert signals from $0\sim10$ GHz (microwave) to the several hundred GHz range (mm-wave). The signals are received and then mixed down to $0\sim10$ GHz so that they can be measured by a microwave Vector Network Analyzer (VNA). Since $\alpha\propto k^3V/Q$ must remain unchanged, the quality factor $Q$ must be the same as the full-scale cavity. For an empty metallic enclosure with loss dominated by ohmic loss in the walls, the quality factor can be estimated as $Q\approx3V/(2S\delta)$ where $S$ is the wall surface area, $\delta=\sqrt{2/(\omega \mu \sigma)}$ is the skin depth in the local limit, and $\sigma$ is the electrical conductivity. After the scaling, setting $Q'=Q$ leads to $\delta'=\delta/s$, and thus $\sigma'=\sigma s$. Conductivity scaling can be achieved by changing the cavity material to a better conductor and by cooling the cavity down to low temperatures using a cryostat.

\section{Experimental Setup}
In our setup, we scale down a 66 cm by 122.5 cm by 127.5 cm aluminum ``full-scale'' cavity designed for the $3.7\sim5.5$ GHz range (WR187 band) by a factor of 20 in each dimension, i.e. $s=20$. The new frequency range becomes $75\sim110$ GHz (WR10 band), which can be measured by using a Keysight network analyzer (KT-N5242A 10 MHz to 26.5 GHz PNA-X ) working together with two VDI frequency extenders (Tx/Rx WR10 module). To achieve higher $Q$, the miniature cavity is made of oxygen-free high-conductivity (OFHC) copper, with mechanically polished inner wall surface to reduce the surface resistance \cite{roughnessPhDThesis,roughnessTechReport}. We then use a custom-built BlueFors BF-XLD400 cryogen-free dilution refrigerator system, which can reach a base temperature of 10 mK under minimum heat-load conditions, to cool the cavity and further increase $Q$. The available volume for samples is a cylinder of 50 cm in diameter and 50 cm in height, that has a total volume of $V\approx(150\lambda)^3$ at 100 GHz, providing abundant space for larger structures. 

\onecolumngrid
\begin{figure*}
\begin{centering}
\includegraphics[width=\textwidth]{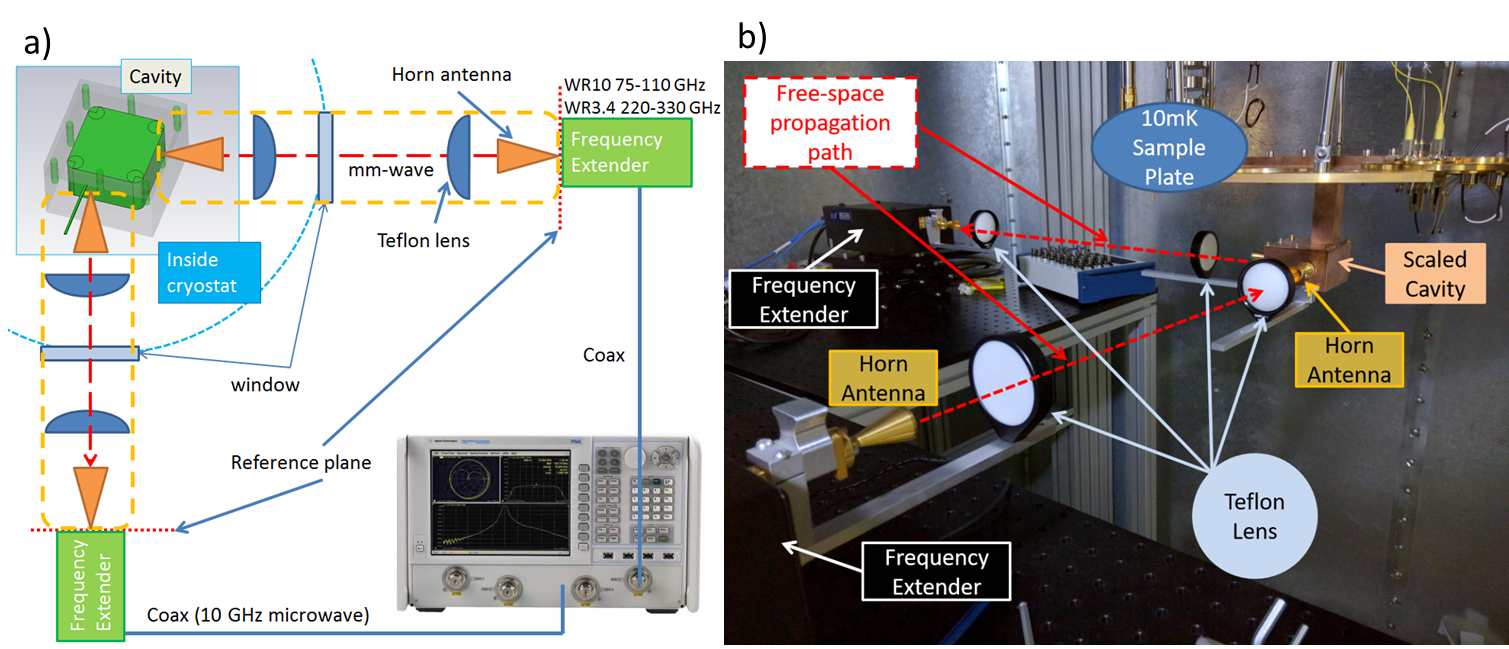}
\caption{\label{fig:path} Schematic diagram and picture of the experimental setup. High frequency waves  propagate in free-space from the frequency extender to the cavity, and from the cavity to the receiving frequency extender. The horn antenna launches the electromagnetic waves into space and the Teflon lens collimates the waves into a parallel beam. The signal then goes through a focusing lens and enters the cavity through a receiving horn antenna. The outgoing waves follow a similar path to reach the second frequency extender. }
\end{centering}
\end{figure*}
\twocolumngrid

Since the miniature cavity is sitting inside the evacuated cryostat at low temperature, it is not possible to employ an input connection from the signal source to the cavity via a coaxial cable or waveguide. Accordingly, we use a quasi-optical free-space propagation path similar to that of a collimated beam in an optical experiment. As shown in Fig.\ref{fig:path}, the high frequency electromagnetic wave emerging from the frequency extender is launched into air by a horn antenna, and then collimated by a teflon lens. The output is a collimated beam propagating in free-space like a plane wave. The receiving end has a focusing lens, identical to the one on the source side, and a receiving horn antenna which is mounted on the wall of the cavity to transmit the received wave into the cavity. Two such free-space propagation paths are used for the two cavity ports, one path for each port. Figure \ref{fig:path} shows the experimental setup highlighting the free-space propagation path, the frequency extenders, the horn antennas, and the lenses.

\begin{figure}
\begin{centering}
\includegraphics[width=0.3\textwidth]{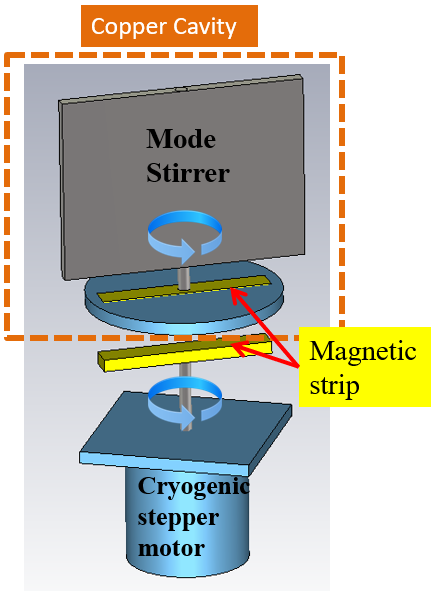}
\caption{\label{fig:motor} Magnetically coupled mode stirrer powered by a cryogenic stepper motor. The magnetic strip outside the cavity (lower yellow bar) is coupled by its static magnetic field to the magnetic strip inside the cavity (upper yellow bar), eliminating the need for any opening on the wall or direct mechanical contact.}
\end{centering}
\end{figure}

Since the RCM is a statistical theory, an ensemble is required to determine the system-specific features and the statistical properties of the enclosures. Consequently, we need to perturb the cavity modes while maintaining the volume of the cavity such that each measurement is a unique realization of the cavity with the same loss parameter. A typical method to create many realizations is to rotate a large metal panel inside the cavity (a ``mode stirrer''), as used in Refs.\cite{zachgigabox,PhysRevE.88.062910,PhysRevLett.110.063902,sameerRCM}. For this purpose, we designed a magnetically coupled mode stirrer powered by a cryogenic stepper motor (Phytron VSS 52.200.2.5‐UHVC suitable for space applications), as shown in Fig.\ref{fig:motor}. The motor rotates a magnetic strip outside the cavity which is magnetically coupled to another magnetic strip inside the cavity, thus eliminating the need for an opening on the wall or direct mechanical contact. The metal mode-stirring panel is attached to the inside magnetic strip and rotates when the stepper motor rotates. In experiments, the motor rotates a small step then waits for the Vector Network Analyser (VNA) to measure the S-parameters of the cavity in the current realization. When the VNA measurement is complete, the motor rotates again, and this process is repeated. In this way data for 200 highly uncorrelated realizations of the cavity is collected and used  to obtain statistics of the electromagnetic properties, and to calculate the ensemble average required by the RCM to characterize system-specific properties. 

\section{Finding The Loss Parameter $\alpha$}
In this section, we discuss two methods for determining the value of the loss parameter $\alpha$ of the enclosure, which governs the statistics of the universal fluctuations: (a) measuring $Q$ and using the direct definition $\alpha=k^3V/(2\pi^2Q)$ and (b) measuring the fluctuations of impedance, and using the RCM normalization process to deduce $\alpha$ by fitting these fluctuations to the prediction of RMT. For a single cavity the loss parameter $\alpha$ uniquely predicts the statistics of the normalized impedance, offering a concise summary of the system statistical properties. However, since we used the remote injection setup shown in Fig.\ref{fig:path}, the data analysis must be  modified to compensate for the extra loss incurred in the free-space propagation path. The comparison between the value of $\alpha$ calculated from $\alpha=k^3V/(2\pi^2Q)$ and that from the modified RCM analysis verifies the validity of the remote injection method.

\subsection{Obtaining $\alpha$ from enclosure $Q$}
\begin{figure*}
\begin{centering}
\includegraphics[width=\textwidth]{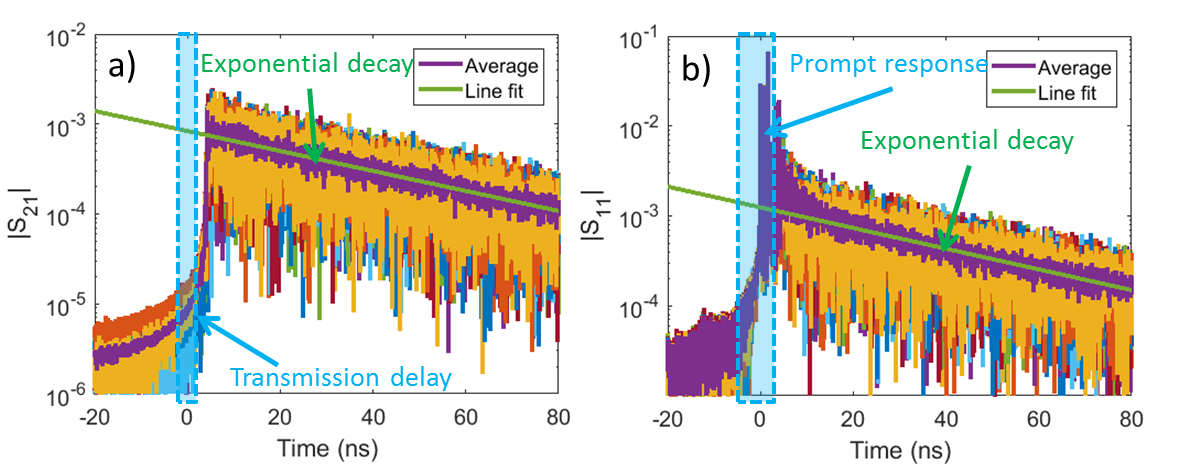}
\caption{\label{fig:etaIfft} The inverse Fourier transform of the measured S-parameters give the value of $\tau$ from each fit in log-scale versus time. a) For the case of transmission and b) for the case of reflection of the $s=20$ scaled enclosure measured through remote injection. Data from 9 realizations at room temperature are plotted with different colors. The purple line is the average, and the green line is the linear fit for the energy decay portion of the average. The slope of the fitted line is $-1/2\tau$, where $\tau$ is the energy decay time of the cavity.}
\end{centering}
\end{figure*}

The quality factor can be calculated according to $Q=\omega \tau$ where $\tau$ is the characteristic energy decay time. To estimate $\tau$ over a given frequency range, we plot the inverse Fourier transform of the measured S-parameters (from 75 GHz to 110 GHz) on a logarithmic scale versus time for an ensemble of 9 realizations, as shown in Fig.\ref{fig:etaIfft}. These plots are equivalent to bandwidth-limited impulse responses in the time domain. The plots for transmission ($|S_{12}|=|S_{21}|$, Fig.\ref{fig:etaIfft} (a)) start with a short delay followed by a exponential decay with a slope of $-1/(2\tau)$. The factor of 2 comes in because $\tau$ is the decay time for energy but the y-axis is proportional to the magnitude of voltage. The plots for reflection ($|S_{11}|$, Fig.\ref{fig:etaIfft} (b)) show an initial prompt response from the antenna, which contains information about the antenna's radiation impedance $Z_{\textnormal{rad}}$ \cite{bisrat,bisratTimeGating}, followed by the same exponential decay. Notice that, even though the 9 curves are somewhat different from each other, their average is very well approximated by a straight line on this log-linear plot. The fluctuations in each curve represent the cavity modes, which are randomly perturbed. Note that this $Q$ is an average over all the modes in the 75 - 110 GHz frequency range.

\onecolumngrid
\twocolumngrid

\begin{figure}
\begin{centering}
\includegraphics[width=0.5\textwidth]{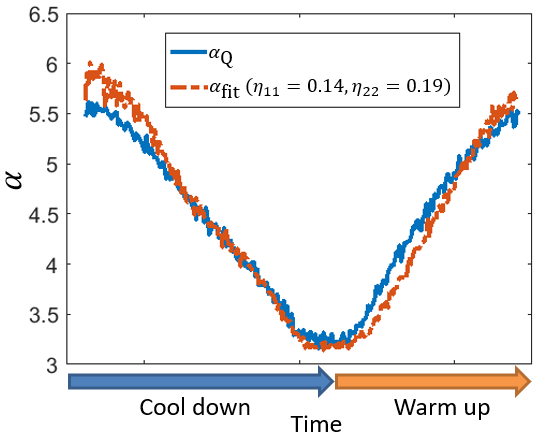}
\caption{\label{fig:alphacompare3} Cycling of the scaled cavity experiment from room temperature to 15 Kelvin and back again, a comparison between resultant $\alpha$ values calculated from different methods. Blue solid line: calculated from time domain energy decay time method; red dotted line: calculated from the best fit of $\eta_{11}$ and $\eta_{22}$ to the RCM prediction at room temperature.}
\end{centering}
\end{figure}

The center frequency for this range used in Fig. \ref{fig:etaIfft} is $f=92.5$ GHz, the cavity volume is $V=1.289\times 10^{-4} \textnormal{ m}^3$, and the quality factor obtained at room temperature from the measured decay time is about $Q=8450$, giving a loss parameter of $\alpha_Q= 5.6$ ($\alpha_Q$ denotes calculated from the quality factor). The same technique is applied to the thousands of S-parameter data sets collected during a cool-down/warm-up temperature cycle, which ranges from room temperature to 15 Kelvin to room temperature, as shown in Fig.\ref{fig:alphacompare3} by the blue solid line. It is seen that by choosing a temperature, we can set the cavity $\alpha$ to any value between 3.3 and 5.6. Note that this determination of $\alpha_Q$ is independent of the loss introduced by the free-space propagation paths.

\subsection{Obtaining $\alpha$ from fits to universal impedance fluctuations}
To calculate $\alpha$ using the RCM normalization process, we first note the following expression for the normalized impedance $\boldsymbol{\xi}$ was obtained in Refs. \cite{zachgigabox,sameerRCM},

\begin{equation}
\label{equ:normalization}
\boldsymbol{\xi}=(\textnormal{Re}[\mathbf{Z}_{\textnormal{avg}}])^{-1/2}(\mathbf{Z}_{\textnormal{cav}}-j \textnormal{Im}[\mathbf{Z}_{\textnormal{avg}}])(\textnormal{Re}[\mathbf{Z}_{\textnormal{avg}}])^{-1/2}
\end{equation}
where $\mathbf{Z}_{\textnormal{cav}}$ is the measured cavity impedance, $\mathbf{Z}_{\textnormal{avg}}$ is the ensemble average of $ \mathbf{Z}_{\textnormal{cav}}$ over many stirrer positions. However, we note that in Refs. \cite{zachgigabox,sameerRCM}, Eq. (\ref{equ:normalization}) was derived assuming lossless ports, hence for our remote injection setup we first need to modify the normalization Eq.(\ref{equ:normalization}) to compensate for the loss during the free-space propagation path. In the case of a \textit{high loss} one-port system with a lossy port (antenna), Ref. \cite{bisrat} shows that 
\begin{equation}
\label{equ:xi1}
\xi=(Z_{\textnormal{cav}}-Z_{\textnormal{avg}})/(\eta\textnormal{Re}\left[Z_{\textnormal{avg}}\right])+1,
\end{equation} 
where $\eta$ is the radiation efficiency of the antenna, $1\ge\eta\ge0$,  defined as the ratio of the power radiated to the power delivered to the antenna. Extending the treatment of highly lossy ports in Ref. \cite{bisrat} to $N$-port systems, we obtain (analogous to Eq. \ref{equ:xi1}), 
\begin{equation}
\begin{aligned}
\label{equ:xi2}
\boldsymbol{\xi}&=\mathbf{R}^{-1/2}(\mathbf{Z_{\textnormal{cav}}}-\mathbf{Z}_{\textnormal{avg}})\mathbf{R}^{-1/2} + \mathbf{I} \\
\mathbf{R}&=\boldsymbol{\eta}^{1/2}\textnormal{Re}[\mathbf{Z}_{\textnormal{avg}}]\boldsymbol{\eta}^{1/2},
\end{aligned}
\end{equation}
where $\mathbf{R}$, $\mathbf{Z_{\textnormal{cav}}}$ and $\mathbf{Z_{\textnormal{avg}}}$ are $N\times N$ matrices, $\mathbf{I}$ is the $N \times N$ identity matrix, and $\boldsymbol{\eta}$ is 
$$
\boldsymbol{\eta}=
\begin{bmatrix}
    \eta_{11} & 0  & \dots  &0 \\
    0 & \eta_{22}  & \dots  & 0 \\
    \vdots & \vdots & \ddots & \vdots \\
    0 & 0 & \dots  & \eta_{NN}
\end{bmatrix}
,
$$
where $\eta_{ii}$ is the radiation efficiency for the $i^{\textnormal{th}}$ port. It is assumed that the $N$-port cavity is in the high-loss limit ($\alpha \gg 1$).

\onecolumngrid
\begin{figure*}
\begin{centering}
\includegraphics[width=\textwidth]{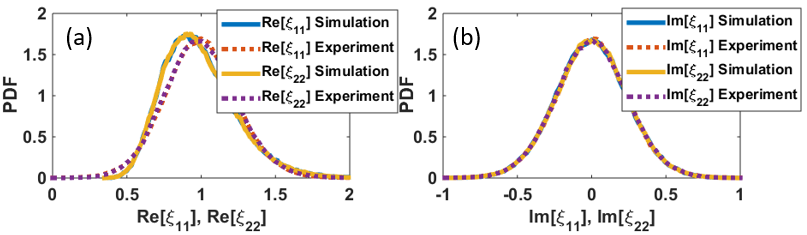}
\caption{\label{fig:etafit} Comparison between the normalized impedance PDF for a 2-port system from a RCM Monte Carlo simulation with $\alpha=5.6$, in solid lines, and that from a normalization process of experimental data with $\eta_{11}=0.14$ and $\eta_{22}=0.19$, in dotted lines.}
\end{centering}
\end{figure*}
\twocolumngrid

In our 2-port setup, the radiation efficiency  $\boldsymbol{\eta} = \left[ \begin{smallmatrix} \eta_{11}&0\\ 0&\eta_{22} \end{smallmatrix} \right]$ cannot be measured directly. Instead, we utilize the knowledge of the loss parameter from the direct definition method $\alpha_Q$ and a fitting process to deduce the radiation efficiency. We already know that the enclosure is characterized by $\alpha_Q=5.6$ at room temperature. Thus, by using an RCM Monte Carlo simulation, we can obtain a prediction for the universally fluctuating impedance PDFs (Eq.
\ref{equ:normalization}) of a 2-port system's normalized impedance with $\alpha=5.6$, as shown in Fig.\ref{fig:etafit} with the solid lines. Then we find the best $\eta_{11}$ and $\eta_{22}$ values such that the normalized impedances, calculated with $\boldsymbol{\eta}$ according to Eq.(\ref{equ:xi2}) using the remote injection experimental data, best approximate the PDFs produced by the RCM simulation results. The best fit values are $\eta_{11}=0.14$, $\eta_{22}=0.19$ and the resulting normalized impedance PDFs are plotted in Fig.\ref{fig:etafit} as dotted lines. We believe that the deviations in $\textnormal{Re}[\xi_{11}]$ and $\textnormal{Re}[\xi_{22}]$ statistics (Fig.\ref{fig:etafit} (a)) are because Eq.(\ref{equ:xi2}) only works for high loss cavities ($\alpha\gg 1$) \cite{bisrat}, and $\alpha=5.6$ in our case is not high enough. Applying $\eta$ changes the variance of $\textnormal{Re}[\xi_{11}]$ and $\textnormal{Re}[\xi_{22}]$, but does not change their peak location. Before applying $\eta$, the fluctuations of $\textnormal{Re}[\xi_{11}]$ and $\textnormal{Re}[\xi_{22}]$  are narrowly centered around 1 and remain so afterwards, deviating from the peak location in the simulation. (Efforts are underway to further generalize the treatment of lossy ports in the RCM to accommodate lower loss cavities.) 

The other solid and dotted lines in Fig.\ref{fig:etafit} (b) lies right on top of each other, as well as the curves for the real and imaginary parts of $\xi_{12}$ and $\xi_{21}$ (omitted in Fig.\ref{fig:etafit} for simplicity), proving that the fitted $\boldsymbol{\eta}$ successfully separates the effects of the lossy free-space path from the cavity losses. We have applied this $\boldsymbol{\eta}$ to all other data sets in the same experiment, assuming that the propagation paths are not perturbed as the temperature varies. The resultant $\alpha_{\textnormal{fit}}$ deduced in this manner is plotted in Fig.\ref{fig:alphacompare3} as the red  dotted line, which agrees well with the $\alpha_Q$ curve calculated from the first method. 

\begin{figure}
\begin{centering}
\includegraphics[width=0.5\textwidth]{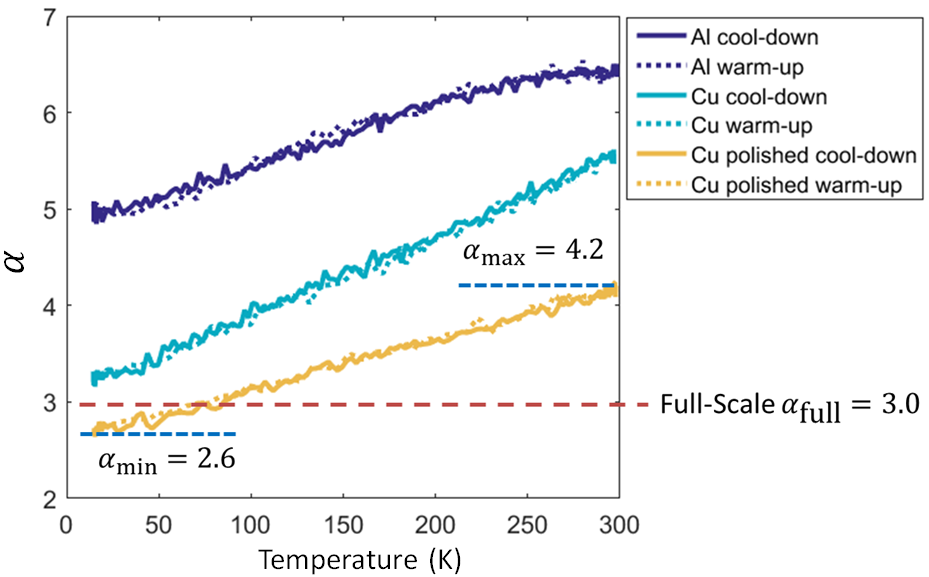}
\caption{\label{fig:alpharange} The tunable range of $\alpha$ values of the $s=20$ scaled cavity using different wall material and varying temperature. The least lossy case is with polished copper walls, and has a range of $2.6\le\alpha\le 4.2$. The overall range is $2.6\le\alpha\le 6.4$.}
\end{centering}
\end{figure}

To maximize the tunable range of the $\alpha$ values, we also vary the cavity wall material in order to vary ohmic loss. We performed the cool-down experiment with the same miniature cavity with three different wall material conditions: copper wall ($\alpha$ results shown in Fig.\ref{fig:alphacompare3}), mechanically polished copper wall, and wall covered with aluminum foil. The polishing reduces the surface roughness and thus reduces surface losses \cite{roughnessPhDThesis,roughnessTechReport}. The overall range of achievable $\alpha$ values are shown in Fig.\ref{fig:alpharange}.

\section{Comparison With Full-Scale Cavity}
\begin{figure}
\begin{centering}
\includegraphics[width=0.5\textwidth]{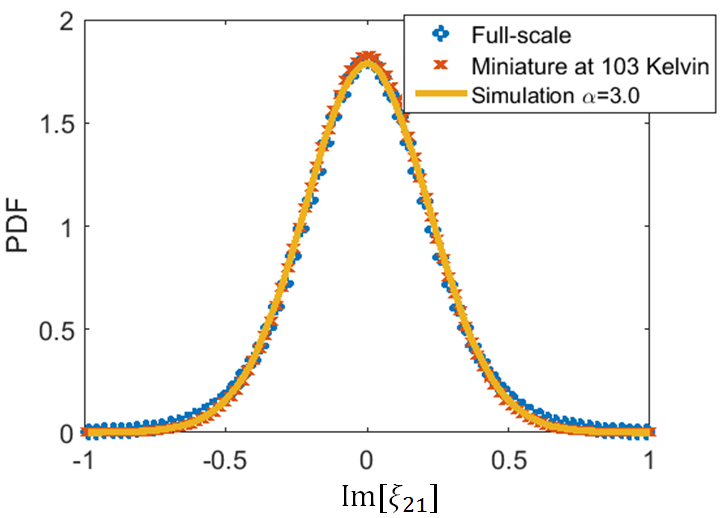}
\caption{\label{fig:fullScaleCompare}  Comparison of the probability density function for the imaginary part of the normalized impedance $\xi_{21}$ for the full-scale cavity (blue diamond dots based on data), the miniature cavity (red cross dots based on data) and the RCM Monte Carlo simulation with $\alpha=3.0$ (yellow solid line) for the entire frequency range either 3.75 - 5.5 GHz or 75 - 110 GHz.}
\end{centering}
\end{figure}

The full-scale cavity is a nearly exact scaled-up version of the miniature cavity, with a scaling of $s=20$ in each dimension. It has an $\alpha$ value of 3.0 in the full scale frequency range (3.75 - 5.5 GHz), which is within the range of the miniaturized cavity's $\alpha$ values in the cool-down experiment (see Fig.\ref{fig:alpharange}). To directly compare the PDF of the normalized impedance, we choose the collected ensemble of data for the miniature cavity with polished copper wall measured around 103 Kelvin, and plot it with the full-scale experimental result, as well as the RCM Monte Carlo simulation result, in Fig.\ref{fig:fullScaleCompare}. In order to show the comparison between the three results, only the imaginary part of $\xi_{21}$ is plotted here, but we analyzed all eight curves (real and imaginary part of $\xi_{11}, \xi_{12}, \xi_{21}$ and $\xi_{22}$) with the same conclusion. We see that all three results agree with each other, confirming that the scaled-down cavity at a particular temperature can reproduce the normalized impedance statistics of the full-scale cavity.

\begin{figure*}
\begin{centering}
\includegraphics[width=\textwidth]{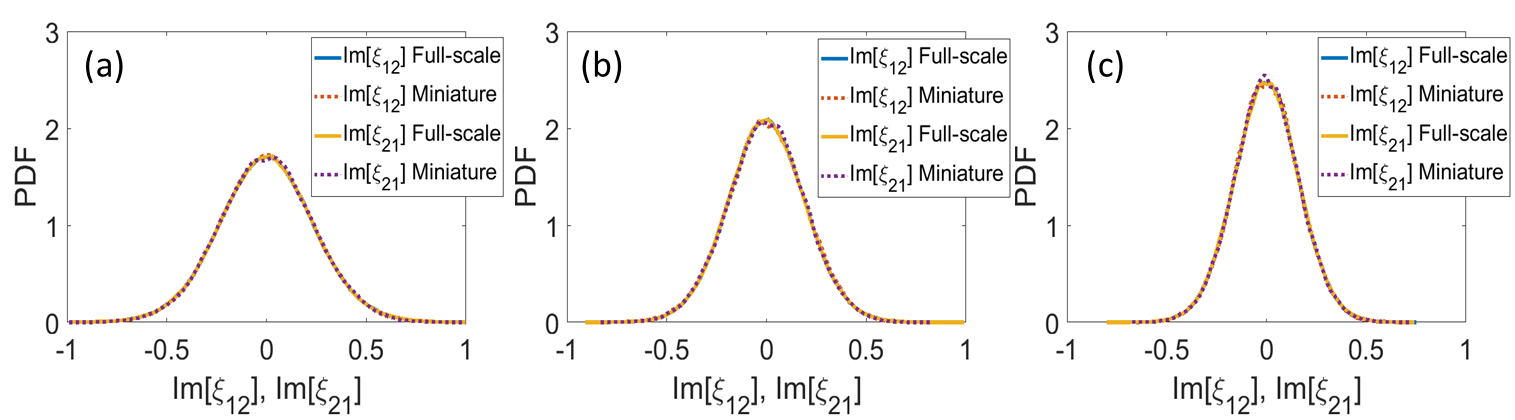}
\caption{\label{fig:alphafreq} Comparison of the probability density function (PDF) for the imaginary part of the normalized impedance $\xi_{12}$ and $\xi_{21}$ between the full-scale cavity (solid line based on data) and the miniature cavity (dotted line based on data) for three different frequency bands (of the full-scale cavity) at different temperatures (of the scaled cavity), (a) $\alpha=2.83$ within [4.45, 4.625] GHz at 130 Kelvin, (b) $\alpha=4.19$ within [4.975, 5.15] GHz at 217 Kelvin and (c) $\alpha=5.82$ within [5.325, 5.5] GHz at 297 Kelvin. Notice that in each plot, all four curves collapse into one because they match each other very well.}
\end{centering}
\end{figure*}

Notice that all analysis presented so far was done using the data for the entire frequency range, (75 - 110) GHz for the miniature cavity and (3.75 - 5.5) GHz for the full-scale cavity. Hence the $\alpha$ values used in the statistics are averaged over a wide frequency range. To see the frequency dependence of $\alpha$, we divide the entire frequency range into 10 sections, 175 MHz wide (3.5 GHz wide in the miniature cavity case) for each section, and carry out the same analysis. Again, we find that we are able to match the normalized impedance statistics by choosing the data recorded at an appropriate temperature such that both cavities had the same $\alpha$; several examples are shown in Fig.\ref{fig:alphafreq} with almost identical curves for full-scale and miniature cavity statistics.

\onecolumngrid
\twocolumngrid

\section{Conclusion}
To summarize, we have presented an experimental setup that scales down a cubic meter microwave cavity, while faithfully maintaining its statistical electromagnetic properties. The setup employs two features to reproduce the correct statistical properties in the scaled cavity. First, it uses frequency extenders to scale up the frequency. Second, it maintains the same wall-loss quality factor by using better electric conductors as the walls of the scaled down cavity and by cooling down the cavity in a cryostat. The experimental results show that the miniature cavity has a wide range of tunable $\alpha$ values from 2.6 to 6.4. We can match the full-scale cavity statistics by choosing the appropriate wall metal and temperature. The agreement is obtained for data selected from a large frequency range, as well as from small frequency sections. These results also demonstrate the capability of characterizing the statistical properties of complex enclosures even under circumstances of remote injection through free space.

\begin{acknowledgments}
This work was supported by ONR under Grant No. N000141512134, AFOSR COE Grant FA9550-15-1-0171 and ONR DURIP grant N000141410772, and the Maryland Center for Nanophysics and Advanced Materials.
\end{acknowledgments}

\section{Appendix}
\subsection{Obtaining loss parameter $\alpha$}
The loss parameter $\alpha$ can be obtained by two methods, depending on what is known about the cavity. If the volume, $V$, and the quality factor, $Q$, is known, then $\alpha$ can be computed directly with its definition $\alpha=k^3V/(2\pi^2Q)$, where $k$ is the wave number. Otherwise one can follow the RCM normalization process below.
\begin{enumerate}
\item First measure the cavity S-parameters, $\mathbf{S}_{\textnormal{cav}}$, using a Vector Network Analyzer (VNA), and convert it to impedance parameters $\mathbf{Z}_{\textnormal{cav}}$ by $\mathbf{Z}=\mathbf{Z}_0^{1/2}(\mathbf{I}+\mathbf{S})(\mathbf{I}-\mathbf{S})^{-1}\mathbf{Z}_0^{1/2}$, where $\mathbf{I}$ is an identity matrix, $\mathbf{Z}_0$ is a diagonal matrix whose elements, $Z_{ii}$, are the characteristic impedances of the transmission line connecting to the $i^{\textnormal{th}}$ port (typically 50 or 75 Ohms for a coaxial cable).
\item Perturb the cavity modes, usually by rotating a large metal panel inside the enclosure, and repeat the measurement for $\mathbf{Z}_{\textnormal{cav}}$, collecting an ensemble of $\mathbf{Z}_{\textnormal{cav}}$ that represent the same cavity statistically.
\item Calculate the ensemble average $\mathbf{Z}_{\textnormal{avg}}=\langle \mathbf{Z}_{\textnormal{cav}} \rangle _{\textnormal{realizations}}$ that summarizes the system specific features such as the radiation impedance and short orbits between the ports, and then normalize $\mathbf{Z}_{\textnormal{cav}}$ by 
\begin{equation}
\boldsymbol{\xi}=(\textnormal{Re}[\mathbf{Z}_{\textnormal{avg}}])^{-1/2}(\mathbf{Z}_{\textnormal{cav}}-j \textnormal{Im}[\mathbf{Z}_{\textnormal{avg}}])(\textnormal{Re}[\mathbf{Z}_{\textnormal{avg}}])^{-1/2}
\end{equation}
\item Comparing the statistics, such as the probability density function (PDF), of the normalized impedance $\boldsymbol{\xi}$ (real and imaginary parts) with the Monte Carlo simulation results with different $\alpha$ values and find the best fit. For an $N$-port system there are $2N^2$ such statistical distributions all of which should be fit simultaneously by single value of $\alpha$. Figure 3 in \cite{Gradoni2014606} is an example of the theoretical predictions for the PDF of normalized impedance for various $\alpha$ values.
\end{enumerate}

This RCM normalization process has been demonstrated to be very effective at removing the system-spacific features, such as the radiation impedance, from the measurement ensemble data, and is a robust method to obtain the loss parameter $\alpha$ for any sufficiently complex enclosure in the highly over-moded regime. 

\subsection{Relationship between loss parameter $\alpha$ and radiation efficiency $\eta$}
As shown in Eq. \ref{equ:xi1}, $\xi=(Z_{\textnormal{cav}}-Z_{\textnormal{avg}})/(\eta\textnormal{Re}\left[Z_{\textnormal{avg}}\right])+1$. Notice that if we define $\delta\xi=\xi-1=(Z_{\textnormal{cav}}-Z_{\textnormal{avg}})/(\eta\textnormal{Re}\left[Z_{\textnormal{avg}}\right])$ and let $\delta\xi_0=(Z_{\textnormal{cav}}-Z_{\textnormal{avg}})/(\textnormal{Re}\left[Z_{\textnormal{avg}}\right])$, then $\delta\xi=\delta\xi_0/\eta$. If the ports are lossy, then $\eta$ provides a simple correction to obtain the universal fluctuations, at least in the high cavity loss case. We can estimate $\alpha$ from the variance of the fluctuating impedance $\xi$ (Appendix B, Method 4 in \cite{sameerthesis}) by 
$$
\begin{aligned}
\alpha&=1/(\pi\sigma^2_{\textnormal{Re}\left[\xi\right]})&=1/(\pi\sigma^2_{\textnormal{Im}\left[\xi\right]}) \\
&=\eta^2/(\pi\sigma^2_{\textnormal{Re}\left[\xi_0\right]})&=\eta^2/\pi\sigma^2_{\textnormal{Im}\left[\xi_0\right]}
\end{aligned}
$$
where $\sigma^2_X$ denotes the variance of $X$. Since $\xi_0$ is independent of the choice of $\alpha$ or $\eta$, its variance is a known constant for a given ensemble. Hence $\alpha/\eta^2=1/(\pi\sigma^2_{\textnormal{Re}\left[\xi_0\right]})=1/\pi\sigma^2_{\textnormal{Im}\left[\xi_0\right]}$ is a constant for a certain data set regardless of the choice of $\eta$. In other words, if $\eta$ is estimated higher than its true value then $\alpha$ will also be higher than it really is. It makes sense that a higher $\eta$, meaning a more efficient and less lossy antenna, leads to a higher $\alpha$, meaning a more lossy cavity, because the total lossyness of the system is fixed for a given ensemble of data.

\bibliography{ref}

\end{document}